\def\draftversion{false}
\def\showall{true}  

\RequirePackage{ifthen}
\ifthenelse{\equal{\draftversion}{true}}{
  \documentclass[aps,pra,10pt,galley,amsmath,amssymb,
                 longbibliography,superscriptaddress]{revtex4-1}
\usepackage{showlabels}
}{
  \documentclass[aps,pra,10pt,twocolumn,amsmath,amssymb,
    longbibliography,superscriptaddress,nofootinbib]{revtex4-1}
}

\usepackage[colorlinks=true,citecolor=blue,linkcolor=blue]{hyperref}

\usepackage{graphicx}
\usepackage[dvipsnames]{xcolor}
\usepackage{amsmath,amsthm,amssymb}
\usepackage{bm,bbm,mathbbol}
\usepackage{oubraces}

\usepackage{soul}

\newcommand{\eq}[1]{Eq.~(\ref{eq:#1})}

\newcommand{\fig}[1]{Fig.~\ref{fig:#1}}

\usepackage{soul}  

\ifthenelse{\equal{\draftversion}{true}}{
  \marginparwidth 2.7in
  \marginparsep 0.5in
  \newcounter{comm} 
  \def\commnext{\stepcounter{comm}}
  \def\commtext{{\bf\color{magenta}[\arabic{comm}]}}
  \def\commmar{{\bf\color{magenta}[\arabic{comm}]}}
  \def\jim#1{\commnext\marginpar{\small JI\commmar: #1}\commtext}
  \def\ism#1{\commnext\marginpar{\small IS\commmar: #1}\commtext}
  \def\fjm#1{\commnext\marginpar{\small FJ\commmar: #1}\commtext}
  
}{
  \def\jim#1{}
  \def\ism#1{}
  \def\fjm#1{}
  
}

\ifthenelse{\equal{\showall}{false}}{
  \renewcommand\includegraphics[2][]{{\bf Figure not shown.}}
  \renewcommand\input[1][]{{\bf Not shown\ }}
}{}

\renewcommand\Im{\operatorname{Im}}
\renewcommand\Re{\operatorname{Re}}

\def\beq{\begin{equation}}
\def\eeq{\end{equation}}
\def\bea{\begin{eqnarray}}
\def\eea{\end{eqnarray}}

\def\ip#1#2{\langle#1\vert#2\rangle}

\def\kk{{\bf k}}

\begin{document}

\title{Directional shift current in mirror-symmetric BC$_2$N}

\author{Julen Iba\~{n}ez-Azpiroz} 
\affiliation{Centro de F{\'i}sica de Materiales, Universidad
  del Pa{\'i}s Vasco (UPV/EHU), 20018 San Sebasti{\'a}n, Spain}

\author{Ivo Souza} 
\affiliation{Centro de F{\'i}sica de Materiales, Universidad
  del Pa{\'i}s Vasco (UPV/EHU), 20018 San Sebasti{\'a}n, Spain}
\affiliation{Ikerbasque Foundation, 48013 Bilbao, Spain}

\author{Fernando de Juan}  \affiliation{Ikerbasque Foundation, 48013 Bilbao, Spain}
\affiliation{Donostia International Physics Center (DIPC), 20018 
San Sebasti\'{a}n, Spain}
\date{\today}

\begin{abstract}
{We  present a  theoretical study of the shift current in a noncentrosymmetric polytype of graphitic BC$_2$N. We find that  the 
photoconductivity {near the fundamental gap}
is strongly anisotropic due to the vanishing of particular tensor components  not foretold by {point-group} symmetry arguments; 
this
is a consequence of dipole selection rules 
imposed by
mirror symmetry, which  imply that the relative parities between valence and conduction bands are key for determining
the directionality of the band-edge response. 
In addition, the band-edge photoconductivity turns out to be rather large, 
with the peak value occurring in an energy range suitable for optical manipulation.
}
\end{abstract}
\maketitle

\section{Introduction}

The conversion of light into electrical current via the photovoltaic effect is one of the most copious supplies of renewable energy on Earth. Traditional solar cells make use of $p$-$n$ junctions to generate a built-in electric field that drives the {photoexcited} electrons. 
{It has long been known that}
single-phase noncentrosymmetric crystals exhibit a different type 
of photovoltaic effect,
{called}
the
\emph{bulk photovoltaic effect}
(BPVE)~\cite{belinicher_photogalvanic_1980,sturman_photovoltaic_1992,fridkin_bulk_2001}.
This is  a nonlinear optical response
{that consists in} the generation of a photovoltage  or
photocurrent  upon light absorption.
In recent
years, the study of the BPVE 
has been reinvigorated by the search for
novel materials with
large photoresponsitivies~\cite{Ma2019,tan-cm16,rangel-prl17,kushnir_ultrafast_2017-ges,kushnir_ultrafast_2019}, as well as by
its sensitivity to geometric and topological properties of the electronic wave functions~\cite{morimoto-sa16,fregoso-prb17,juan-nc17}.
As an example, the sign of the shift-current contribution to the BPVE has been suggested as a possible probe for detecting topological phase transitions~\cite{PhysRevLett.116.237402,yan_precise_2018}.
In addition, recent experimental work on 
TaAs found large and anisotropic 
contributions to the shift-current BPVE
driven by the low-energy Weyl-node physics~\cite{osterhoudt-arxiv17}.

In this work, we report a distinctive shift-current
response at the band edge of a noncentrosymmetric polytype of 
graphitic
BC$_2$N, a layered semiconductor made of
alternating zigzag chains of carbon and boron nitride~\cite{LWC89,PhysRevLett.77.187,PhysRevLett.83.2406,WSI96,WIM95,WIM96,Nozaki96}. 
Our {\it ab initio} calculations 
show that {near the fundamental band gap} the calculated response exhibits  strong anisotropy,
due to the vanishing of certain  tensor components 
not foretold by phenomenological symmetry arguments.
We trace the origin of this anisotropy to 
the mirror symmetry of the crystal, which imposes
selection rules on dipole transitions between 
the band edges.
We capture the essential physics of this phenomenon with a two-band 
${\bf k}\cdot {\bf p}$
model, 
thus providing a suitable framework for a broad class of materials.

 \section{Shift-current BPVE}
 
The BPVE is a nonlinear optical response of the form~\cite{belinicher_photogalvanic_1980}
\begin{align}
J_a = {\sum_{b,c}}\,\sigma^{abc}(\omega) E_b(\omega) E_c(-\omega),
\label{eq:BPVE}
\end{align}
where $a$, $b$, and $c$ are Cartesian indices. 
Since both {the electric field} $\textbf{E}$ and {the current} $\textbf{J}$ 
are odd under inversion, 
the BPVE can only occur in 
systems where this symmetry is broken. 
The right-hand side of \eq{BPVE} can be split into
symmetric and antisymmetric parts under $b\leftrightarrow c$, known respectively as the {\it linear} and {\it circular} BPVE~\cite{belinicher_photogalvanic_1980}; {the former occurs in
piezoelectric crystals, and the latter in gyrotropic crystals.}
The shift current
is the intrinsic (interband) contribution to the linear BPVE.

In the independent-particle picture, the
shift photoconductivity 
takes the form~\cite{sipe-prb00}
\beq
\label{eq:sigma}
\sigma^{abc}(\omega)=
C\int\frac{d^3k}{(2\pi)^3}\sum_{n,m}\,f_{nm}
I^{abc}_{mn}\delta(\omega_{nm}-\omega)\,,
\eeq
where
$C=-\pi g_s|e|^3/2\hbar^2$
is a combination of fundamental constants ($g_s=2$ accounts for 
spin degeneracy),  $f_{nm}=f_{n}-f_{m}$ and
$\hbar\omega_{nm}=E_{n}-E_{m}$ are differences in band
occupations and energies, {and} 
the matrix element {reads}
\beq\label{eq:Imn}
I^{abc}_{mn}={\rm Im}\left(r^b_{mn}r^{c;a}_{nm}\right) +
b \leftrightarrow c.
\eeq
Here $r^b_{mn}$ is the interband dipole 
(the off-diagonal 
part of the Berry connection
matrix ${ A}^{b}_{mn}=i\ip{u_{m\kk}}{\partial_{k_{b}} u_{n\kk}}$) 
which only depends  on the initial and final states $m$ and $n$,
and $r_{nm}^{c;a}$ is a 
generalized derivative defined as
$r^{c;a}_{ nm}=\partial_{k_{a}} r^c_{ nm}
-i\left(A^a_{ nn}-A^a_{ mm}\right)r^c_{ nm}$,
which depends implicitly on {intermediate} virtual states~\cite{sipe-prb00}.

\section{Mirror symmetry and the role of band-edge parities}
We next analyze the constraints imposed  on the shift photoconductivity tensor by the presence of mirror symmetry; 
without loss of generality, we 
choose $M_x: x \rightarrow -x$.

{Let us start with the conventional symmetry analysis of response tensors~\cite{nye-book57,newnham-book05}.}
If $M_x$ leaves the crystal structure invariant,
then according to \eq{BPVE} the tensor $\sigma^{abc}$ can be nonzero only when $x$ appears  an even number of times (zero or two) in $abc$. This is a phenomenological constraint,
which holds at any frequency and 
irrespective of the mechanism behind the BPVE; it also holds
if the space-group operation is not a pure reflection but a glide{, in which case the operation $M_z$ is still present in the point group. More generally, $\sigma^{abc}=\sigma^{acb}$ transforms under point-group operations in the same way as the piezoelectric tensor $d^{abc}=d^{acb}$, whose symmetry-allowed components have been tabulated for every crystal class~\cite{nye-book57}}.

Now assume that the space-group operation is a pure reflection $M_x$, 
and that  the minimum direct gap $E_g=E_c-E_v$ is located on a $M_x$-invariant plane in the BZ. Under these conditions, the shift photoconductivity at frequencies close to $E_g$ is further restricted by dipole selection rules, in much the same way as 
the optical absorption~\cite{bassani_band_1967,bassani_electronic_1975}.
The reason is that the states $\vert v\rangle$ and $\vert c\rangle$ at the top of the valence and  bottom of the conduction band
are now eigenstates of $M_x$, with eigenvalues $\pm i$
in the spinful case and $\pm 1$ in the spinless case; introducing
a ``relative mirror parity'' ${\mathcal P}_{vc}^x$ that equals $+1$ ($-1$) when $\vert v\rangle$ and $\vert c\rangle$ have equal (opposite) eigenvalues, the dipole matrix elements are found to satisfy~\cite{bassani_band_1967,bassani_electronic_1975}
\begin{subequations}
\label{eq:px}
\begin{align}\label{eq:px+1}
r_{vc}^x=0,&\text{ when }{\mathcal P}_{vc}^x=+1,\\
\label{eq:px-1}
r_{vc}^y=r_{vc}^z=0,&\text{ when }{\mathcal P}_{vc}^x=-1.
\end{align}
\end{subequations}
By virtue of Eqs.~\eqref{eq:sigma} and~\eqref{eq:Imn}, these selection rules set some components of $\sigma^{abc}{(\omega)}$ to zero for $\hbar\omega\approx E_g$.

Let us first consider the shift current induced by light 
linearly polarized  along $b=c$. In this case, the phenomenological constraint mentioned above becomes $\sigma^{xbb}=0$, hence the current must 
flow parallel to the mirror plane.
Concerning the band-edge response, 
one can distinguish two scenarios on the basis of \eq{px}. 
When ${\mathcal P}_{vc}^x=+1$
the matrix element $I^{abb}_{mn}$ 
vanishes for $b=x$ so that $\sigma^{axx}=0$,
and when ${\mathcal P}_{vc}^x=-1$ 
it vanishes for  $b\not= x$ so that $\sigma^{ayy}=\sigma^{azz}=0$.
Thus, when ${\mathcal P}_{vc}^x=+1$ (${\mathcal P}_{vc}^x=-1$)
the shift current flows {along the mirror plane}
in response to the component of the optical electric field {that is parallel (perpendicular) to that plane.}

To complete the present discussion, let us consider possible 
contributions to the shift current {from $\sigma^{abc}$} with $b\not=c$. These average to zero for unpolarized light
(\textit{e.g.}, sunlight), 
hence they
are often ignored when discussing
solar-cell applications of the BPVE~\cite{tan-cm16,cook-nc17,wang-prb17}. When they are present, the current is  
no longer constrained to flow along the mirror plane. At the band edge, 
${\mathcal P}_{vc}^x=+1$ imposes no restriction at all,
while ${\mathcal P}_{vc}^x=-1$
forces $\sigma^{abc}$ to vanish if {both $b\neq x$ and $c \neq x$}.

\section{Physical realization}

\begin{figure}[t]
\includegraphics[width=\linewidth]{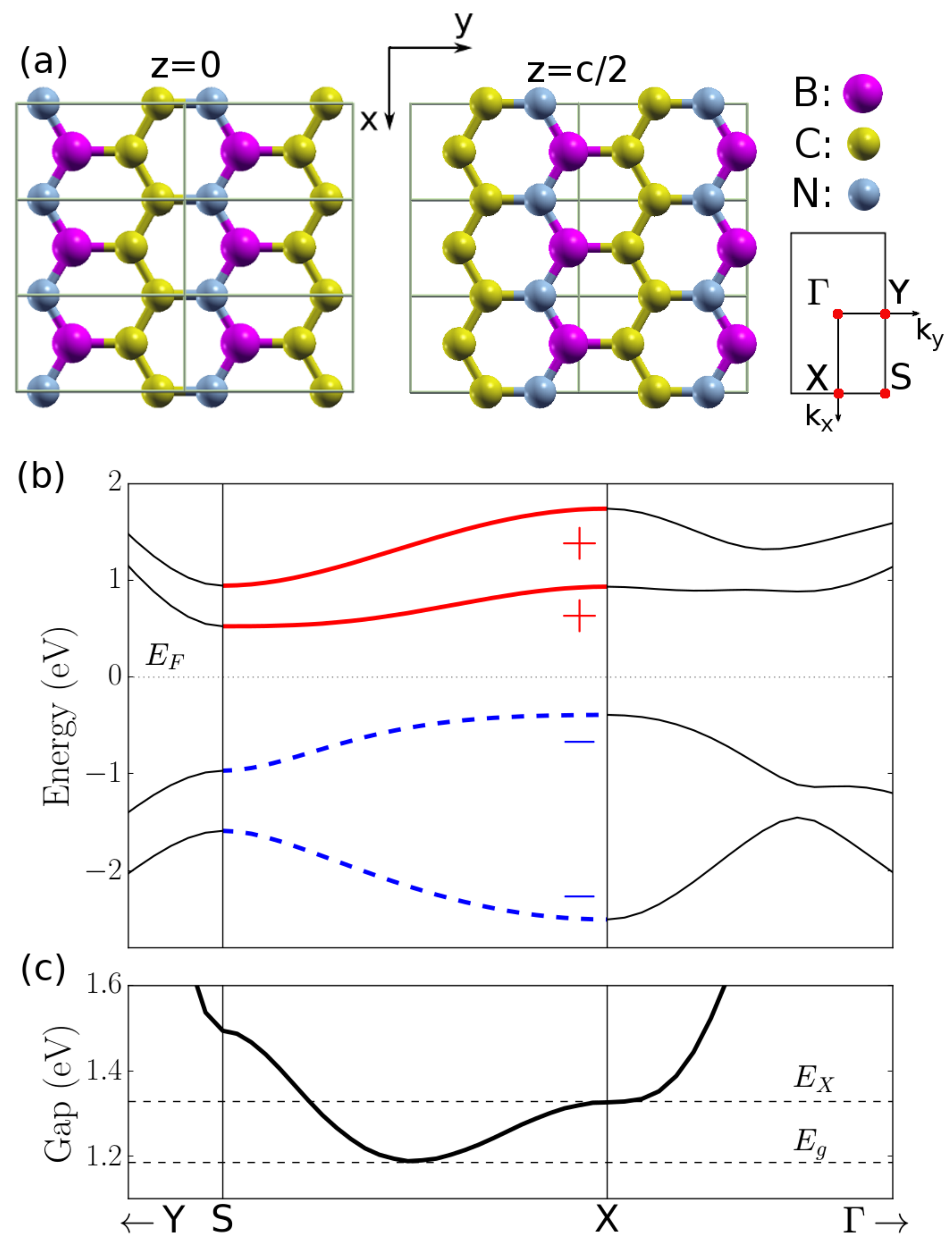}
\caption{(a) Two adjacent layers of the BC$_{2}$N-A2
  structure, with two formula units per cell {of height $c$}.  Solid gray lines delimit the cells (horizontal lines coincide with one family of $M_x$ symmetry planes). The 2D BZ is also illustrated. 
    (b) {Band structure on the $k_z=0$} plane.
    Along S--X, thick solid (red) and dashed (blue) lines denote bands with {$M_{x}$ eigenvalues $+1$ and $-1$,} respectively. Along other symmetry lines 
    bands are drawn as thin solid lines.
    (c) Direct band gap, with values $E_{g}$ and $E_X$ at the band edge and X, respectively. 
    }
\label{fig:1}
\end{figure}

We now show that graphitic BC$_{2}$N provides a striking
illustration of the preceding discussion.
We begin by noting that while a single layer breaks inversion symmetry 
[see  \fig{1}{(a)}],  
whether inversion is still broken in the bulk structure
depends on the stacking pattern, which remains to be determined experimentally.
There are two types of stackings, denoted A or B depending on whether consecutive layers
have the same or opposite orientation~\cite{PhysRevB.73.193304}.
B-type structures have a center of inversion between the layers while those of type~A break inversion symmetry, hence the photoconductivity remains finite only in A-type structures.

We consider the most stable
A-type bulk structure identified in Ref.~\onlinecite{PhysRevB.73.193304}, namely the A2 
structure illustrated in \fig{1}{(a)}.
The space group  is \textit{Pmm2} (No. 25), and the
point group is \textit{mm2}.
There are
two mirrors $M_x$ and $M_z$, 
and a rotation $C_2^y$ about the polar axis.
Point-group symmetry allows 5 out of the 18 independent components
of the linear BPVE tensor $\sigma^{abc}=\sigma^{acb}$ to be nonzero {(the same as for the piezoelectric tensor~\cite{nye-book57})}: 
three involving in-plane directions only 
($yxx$, $xxy=xyx$, and $yyy$), 
and two that also involve $z$ 
($yzz$ and $zzy=zyz$).  
Since $M_x$ and $M_z$ are pure reflections, at the band edge further restrictions emerge from
dipole selection rules
as detailed in Table~\ref{table}.

\begin{table}[t]
\caption{\label{table}
Selection rules for the band-edge shift photoconductivity $\sigma^{abc}(\hbar\omega\approx E_g)$ in the crystal class {\it mm2}. 
Each column lists one of the symmetry-allowed components, followed by the relative $M_x$ and $M_z$ 
band-edge parities
that 
allow that component to be non-negligible 
for $\hbar\omega\approx E_g$. The relative parities for BC$_{2}$N-A2, ${\mathcal P}_{vc}^x=-1$ and ${\mathcal P}_{vc}^z=+1$, are marked in bold; they imply that only $\sigma^{yxx}$ and $\sigma^{xxy}{=\sigma^{xyx}}$ (also marked in
    bold) are non-negligible at the band edge.
    }
    \begin{ruledtabular}
 \begin{tabular}{  c  c   c  c   c  c }
  & \multicolumn{5}{c}{Components of $\sigma^{abc}(\hbar\omega\approx E_g)$}   \\
      &  \textbf{\textit{yxx}} & \textbf{\textit{xxy=xyx}} & \textit{yyy} & \textit{yzz} & \textit{zzy=zyz}        \\
  \hline
${\mathcal P}_{vc}^x$  & $\bm{-1}$   & +1/$\bm{-1}$ & +1          & +1   & +1    \\
${\mathcal P}_{vc}^z$  & \textbf{+1} & \textbf{+1}  & \textbf{+1} & $-1$ & \textbf{+1}/$-1$  \\
\end{tabular}
\end{ruledtabular}
\end{table}

The scalar-relativistic
band structure of BC$_{2}$N-A2 
near the band edge
is displayed in \fig{1}{(b)}. We only show the 
dispersion {on the $k_z=0$ plane},
because the weak interlayer 
coupling produces a quasi-2D band structure with
virtually no $k_z$ dispersion~\cite{PhysRevB.73.193304}. 
Inspection of the figure reveals that 
the dispersion from~S to~X is also relatively weak. The minimum direct band gap of
$E_g\approx 1.18$~eV is located approximately midway between
those two time-reversal invariant momenta (TRIM), as shown in \fig{1}(c).
On the S--X line, whose points remain invariant under $M_{x}$, the
energy eigenstates are also eigenstates of $M_{x}$, with eigenvalues
$\pm 1$ as depicted by the solid and dashed lines in \fig{1}{(b)}. 
We have explicitly verified~\cite{irrep} that the 
upper-valence and lower-conduction bands have
opposite $M_x$ eigenvalues, {\it i.e.}, ${\mathcal P}_{vc}^x=-1$.
Moreover, since all bands in \fig{1}{(b)} are derived
from $p_z$-type Wannier functions (see below), they all have {the same} $M_z$ eigenvalue $-1$ on the $k_z=0$ plane, hence ${\mathcal P}_{vc}^z=+1$. 
With these two parity values in hand, we predict from Table~\ref{table} that the components of $\sigma^{abc}$ that are expected to be present at the band edge are $yxx$ and $xxy=xyx$.

We have computed the shift photoconductivity of BC$_{2}$N-A2 by means of density-functional theory (DFT), using he {\tt Quantum ESPRESSO} code package~\cite{gianozzi-jpcm09}. We took the
structural parameters from Ref.~\onlinecite{PhysRevB.73.193304}, and used scalar-relativistic
pseudopotentials with the Perdew-Burke-Ernzerhof exchange-correlation (XC)
functional~\cite{perdew-prl96}. 
Maximally-localized Wannier functions~\cite{marzari-prb97,souza-prb01} 
were generated with the {\tt
  Wannier90} package~\cite{wannier90}, starting from
atom-centered $p_z$ orbitals for
modelling the bands around the Fermi level.
Finally, the photoconductivity
was calculated using a recently-developed Wannier-interpolation scheme~\cite{PhysRevB.97.245143}.
 
\begin{figure}[t]
\includegraphics[width=\columnwidth]{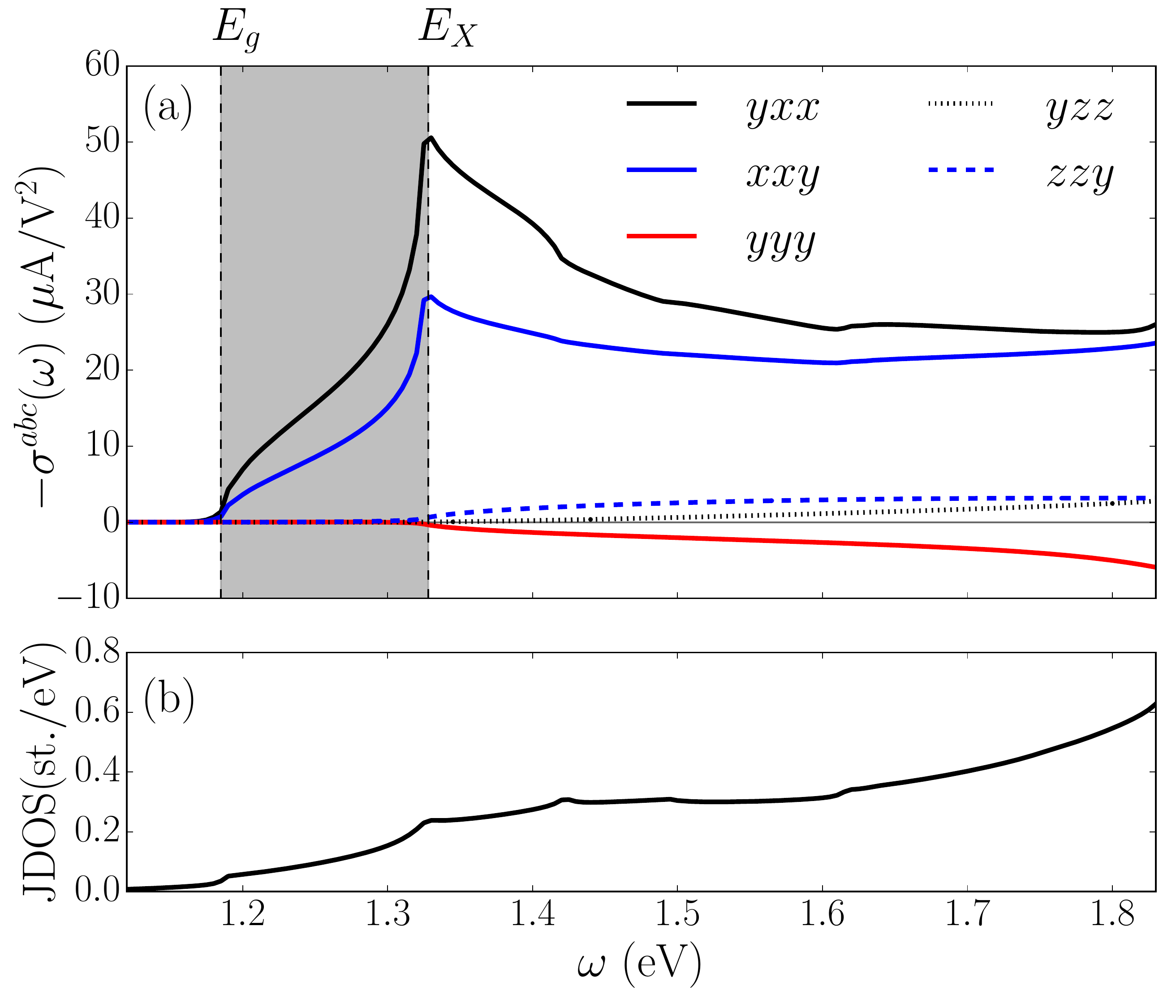}
\caption{(a) Shift photoconductivity of bulk BC$_{2}$N-A2 calculated {\it ab initio}. 
  The gray area indicates the band-edge energy range from $E_g$  to $E_{X}$ [see \fig{1}(c)]. (b) Calculated JDOS. }
\label{fig:2}
\end{figure}

The calculated photoconductivity is shown in \fig{2}(a). As predicted in Table~\ref{table}, three of the five independent components {that are in principle allowed by point-group symmetry}, $\sigma^{yyy}$, $\sigma^{zzy}$ and $\sigma^{yzz}$, have negligible values in the band-edge region indicated by the gray area. The other two, $\sigma^{yxx}$ and $\sigma^{xxy}$, grow rapidly from the onset at $E_g$ until reaching peak values of $\sigma^{yxx}\sim50~\mu\text{AV}^{-2}$  and $\sigma^{xxy}\sim30~\mu\text{AV}^{-2}$ at $E_X\approx 1.33$~eV;
above $E_X$ they drop gradually and then stabilize at roughly half their peak values, before peaking again near 2 eV (not
shown).
The {three} previously-negligible components become sizeable above $E_X$ due to contributions from valence and conduction bands outside the mirror-invariant S--X line [see Figs.~\ref{fig:1}(b,c)], but they remain small compared to the other two.

For light with linear polarization  along a crystallographic axis 
[$b=c$ in \eq{BPVE}], the spectrum in \fig{2}(a) can be rationalized as follows;
at all frequencies $\hbar\omega>E_g$ the shift current flows along the line of intersection between the two mirror planes (along $y$), and at band-edge frequencies it only flows in response to the field component along~$x$ 
(normal to $M_x$  with $P_{vc}^x=-1$, and parallel to  $M_z$  with $P_{vc}^z=+1$).

We now {turn our attention to} the strong peak in $\sigma^{yxx}$ at $E_X$. 
The peak value is
rather large for a gapped {bulk} material; 
{for comparison, among the largest values reported in the literature are}
$\sim 50~\mu\text{AV}^{-2}$ in ferroelectric PbTiO$_{3}$ at $6$~eV~\cite{young-prl12},
and $\sim 80~\mu\text{AV}^{-2}$ at 1.3 eV in the chiral 
crystal RhBiS~\cite{PhysRevB.100.245206}. 
The peak photoconductivity
in BC$_2$N 
occurs at a frequency of $\hbar\omega\approx 1.3$~eV that is suitable for optical manipulation, and where bulk semiconductors typically have much smaller responses~\cite{nnano-bife03-2010,ji-prb-2011,tan-cm16}.
The characteristics of representative photovoltaic materials
are collected in Table~\ref{table:magnitude}, where the last two entries 
correspond to
low-dimensional materials; while their peak photoconductivities
surpass that of BC$_{2}$N-A2, the reported values depend on an
adjustable interlayer-distance
parameter~\cite{rangel-prl17,cook-nc17}. Furthermore, the 3D
crystallization of these low-dimensional structures may not occur in
reality, or it may restore inversion symmetry~\cite{centrosym1986}.
{Free from these concerns, BC$_2$N is a truly bulk material with a
large and highly directional photoconductivity in the visible range.}

\begin{table}[t]
\caption{\label{table:magnitude}
{
Peak shift photoconductivities, peak frequencies and 
employed XC functionals, for a collection of bulk and low-dimensional photovoltaic materials.}
    }
    \begin{ruledtabular}
{
 \begin{tabular}{  c  c   c  c }
  &$\sigma^{abc}$ ($\mu\text{AV}^{-2}$) & $\omega$ (eV) & XC functional \\
    \hline
     BC$_{2}$N-A2 (this work) & 50 & 1.3 & GGA \\
     PbTiO$_{3}$, BaTiO$_{3}$~\cite{young-prl12} & $50$, 30 & 6.0, 6.5 & GGA \\
     GaAs~\cite{nastos-prb06} & 40 & 5.5 & LDA + sciss. \\ 
      LiAsSe$_2$, NaAsSe$_2$~\cite{BYZ14}  & 13, 15 & 2.0, 3.1 & GGA + sciss. \\ 
      BiFeO$_3$~\cite{YZR12}  & 0.8 & 3.5 & GGA + U\\
      RhBiS, IrBiSe~\cite{PhysRevB.100.245206} & 80, 40 & 1.3, 2.1 & GGA\\
      CaAlSiH~\cite{brehm_predicted_2018}& 6 & 1.3 & GGA \\
      \hline
      2D GeS, GeSe~\cite{rangel-prl17}  & 160, 200 & 2.8, 2.0 & GGA + sciss.\\
      1D polymers~\cite{liu_giant_2017}  & 60$-$180 & 0.6$-$0.8 & GGA \\ 
                \end{tabular}
          }
\end{ruledtabular}
\end{table}

{
In order to 
make a more direct connection with potential experimental measurements, it is
useful to report two other figures of merit in addition to the photoconductivity:
the \emph{Glass coefficient}~\cite{glass-apl74,tan-cm16},
which quantifies photocurrent generation in bulk materials 
taking absorption into account, and the 
\emph{shift distance}~\cite{nastos-prb06} (or \emph{anisotropy distance}~\cite{Baltz1981}),
which gives an estimate of 
the real-space shift undergone by an electron upon
photoexcitation.
For light linearly polarized along $b$, the Glass coefficient
is defined as
\beq
G^{abb} =  \dfrac{1}{2c\epsilon_{0}} \dfrac{\sigma^{abb}(\omega)}{\alpha^{bb}(\omega)},
\eeq
with $\epsilon_0$ the vacuum permittivity, 
$\alpha^{bb}(\omega)=2\omega\Im \sqrt{\epsilon^{bb}(\omega)}/c$ the absorption coefficient,
and $\epsilon^{bb}(\omega)$ the {complex} dielectric function.
In turn, the shift distance is defined as~\cite{nastos-prb06}
\beq
d^{abb} = \dfrac{\hbar}{e}\dfrac{\sigma^{abb}(\omega)}{\Im \epsilon^{bb}(\omega) }.
\eeq
At frequencies near the band-edge 
where $\Im \epsilon^{bb}(\omega)\ll\Re \epsilon^{bb}(\omega)$,  we have
$\Im \sqrt{\epsilon^{bb}(\omega)}\simeq \Im \epsilon^{bb}(\omega)/2$, leading to the relation
\beq\label{eq:glass-shiftdist}
G^{abb} \simeq \dfrac{e}{2\epsilon_0} \dfrac{d^{abb}}{\hbar\omega}.
\eeq
}

\begin{figure}[t]
\includegraphics[width=1.0\columnwidth]{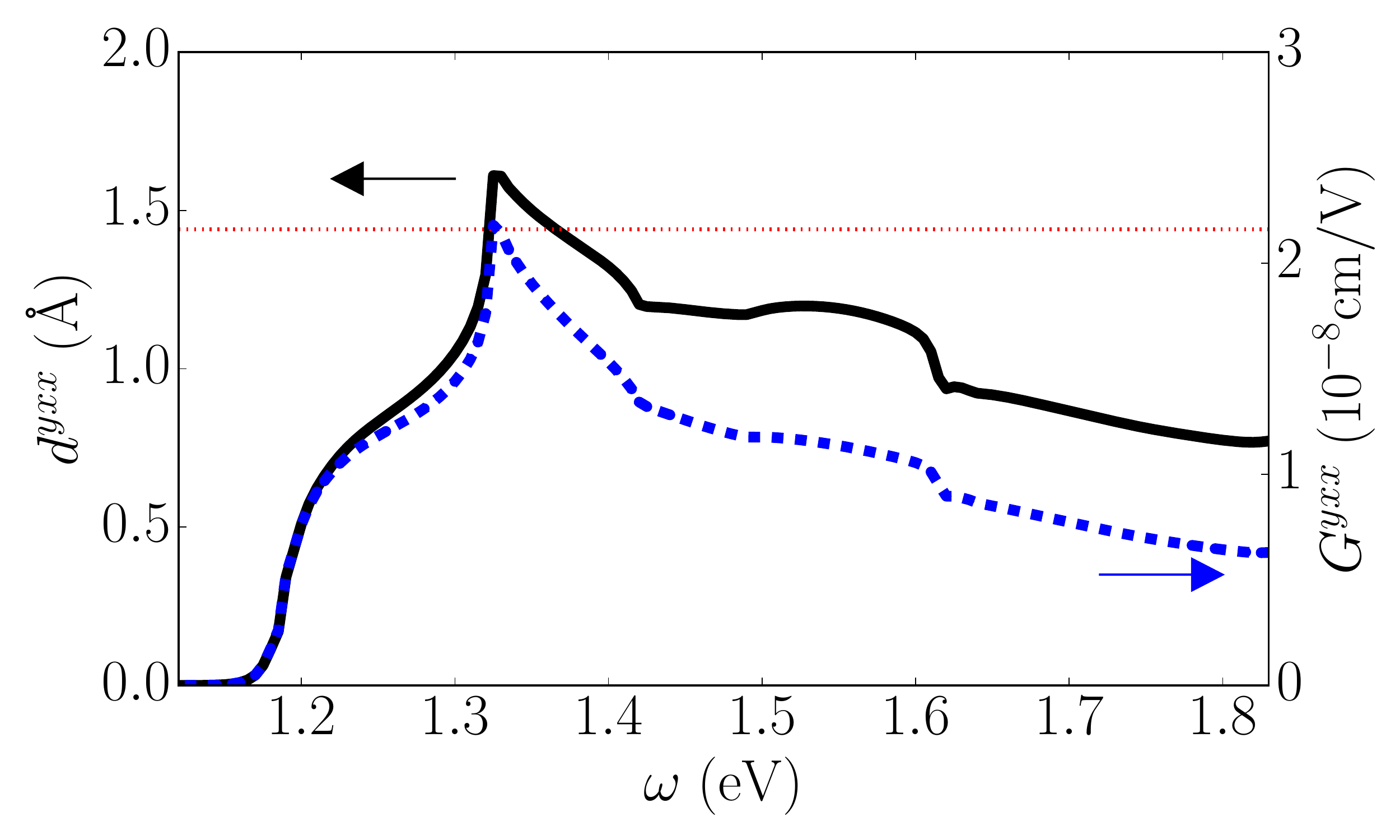}
\caption{{Solid (black) and dashed (blue) lines show the $yxx$ component of the 
shift distance and Glass coefficient whose corresponding ordinate axes are placed on the
left (black) and right (blue) of the graph, respectively. 
The horizontal dotted (red) line denotes the average bond length 
between C--N and C--B distances.}
    }
\label{fig:3}
\end{figure}

{
The quantities $d^{yxx}(\omega)$ and $G^{yxx}(\omega)$ are plotted in \fig{3}. 
Both {spectra} are similar to the
photoconductivity spectrum in \fig{2}(a), with prominent peaks at $~1.3$ eV. The maximum 
shift distance is slightly
larger than the average bond length {between C--N and C--B distances}
indicated by the dashed horizontal line at $1.44$~\AA\,.
This means that despite the large shift-current photoconductivity, 
the real-space shift 
of the photoelectron is not particularly large.
On the other hand, the peak {value of the} Glass coefficient  is $2\cdot 10^{-8}$ cm/V, which ranks 
among the largest reported {to date}~\cite{osterhoudt-arxiv17}.
This difference in the relative magnitudes of the two quantities 
is partly due to the $1/\omega$ factor in 
\eq{glass-shiftdist}.
}

\begin{figure}[b]
\includegraphics[width=0.95\columnwidth]{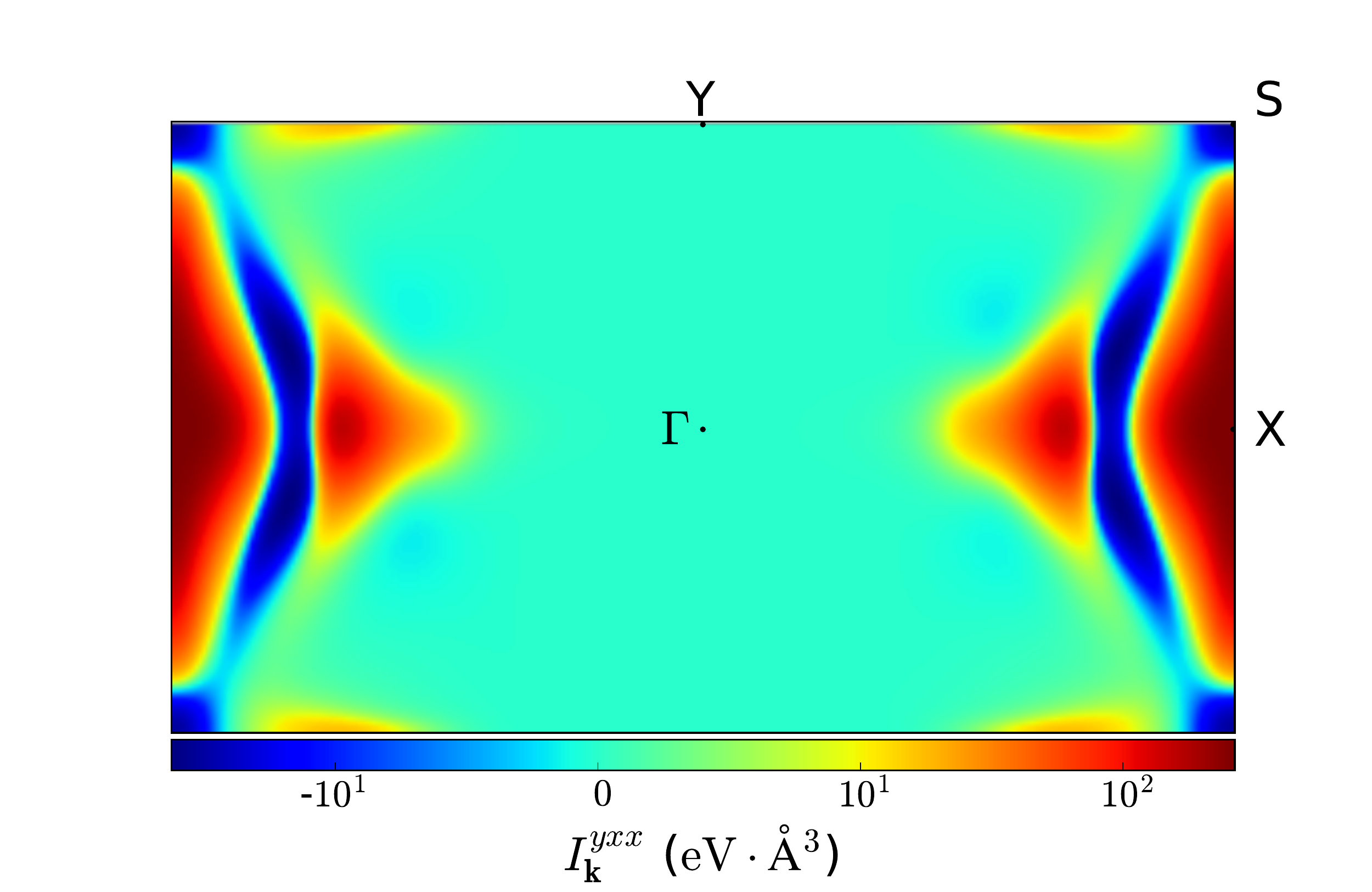}
\caption{Heatmap plot across the 2D BZ at $k_z=0$ of the matrix element
  $I^{yxx}_{\textbf{k}}=\sum_{m,n}^{v,c}I^{yxx}_{mn\kk}$.
    }
\label{fig:4}
\end{figure}

To analyze the photoconductivity near the band edge, we approximate it as the product between the shift-current matrix element and the JDOS~\cite{cook-nc17},
\begin{equation}\label{eq:approx-sigma}
\sigma^{abc}(\omega) \simeq C \;   I^{abc}_{vc}(\omega) N(\omega). 
\end{equation}
The JDOS, plotted in \fig{2}(b), exhibits a 2D-like Van Hove singularity at {$E_X$} due to a   saddle point in the direct band gap [see \fig{1}(c)], boosting the electronic transitions that contribute to the shift current around that energy.  Moreover, those transitions carry a sharply-enhanced matrix element for $\sigma^{yxx}$. This can be seen in \fig{4}, which displays a heatmap plot of the matrix element $I^{yxx}_{mn\kk}$ of \eq{sigma} summed over the upper-valence and lower-conduction bands: 
around X, 
it is {positive-valued and} 
more than two orders of magnitude larger than almost anywhere else in the BZ.

\section{Two-band model in 2D}

Motivated by the quasi-2D nature of graphitic BC$_2$N-A2,
we now construct a minimal 2D model 
that captures the 
mirror-parity effect on the 
photoconductivity near 
the band edge.
The model lies flat on the $(x,y)$ plane, and has both $M_x$ and $M_z$ 
symmetry. For simplicity,
we assume that the band edge lies
at a TRIM on a $M_{x}$-invariant line in the 2D BZ.
With these constraints,  
the most general two-band ${\bf k}\cdot{\bf p}$ Hamiltonian 
that can be obtained by expanding 
up to second order in ${\bf k}{=(k_x,k_y)}$ around the TRIM is 
\begin{equation}
\begin{split}
&H{(k_x,k_y)} = (\alpha_x k_x^2 + \alpha_y k_y^2+ \alpha_{xy}k_x k_y)\sigma_x  +\\  &(v_x k_x+v_y k_y)\sigma_y + (\Delta + \beta_x k_x^2 + \beta_y k_y^2+\beta_{xy}k_x k_y)\sigma_z\,.
\label{eq:H2D}
\end{split}
\end{equation}
Essentially the same model was considered in Ref.~\onlinecite{cook-nc17}, with the following differences. 
(i) We chose $M_{x}$ as the vertical mirror, which requires keeping terms linear in $k_x$ and $k_y$. (ii) 
We took as basis states the energy eigenstates at the valence and conduction band edges, making our Hamiltonian
diagonal at $\textbf{k} =0$.

{Since our basis states are also eigenstates of $M_{x}$ with eigenvalues $\pm 1$, the operator $M_x$ is represented by the identity matrix when 
${\mathcal P}_{vc}^x=+1$ and by $\sigma_z$ when 
${\mathcal P}_{vc}^x=-1$.
Applying $M_xH({\bf k})M_x^{-1}=H(M_x{\bf k})$  to \eq{H2D},
we find
\begin{subequations}
\label{eq:even-odd}
\begin{align}
\label{eq:even}
v_{x}=\alpha_{xy}=\beta_{xy} = 0,\quad
&\text{when $P_{vc}^x=+1$},\\
\label{eq:odd}
\alpha_x = \alpha_y = v_y = \beta_{xy} = 0,\quad
&\text{when $P_{vc}^x=-1$}.
\end{align}
\end{subequations}
The relative band-edge parity ${\cal P}_{vc}^x$ therefore defines two  classes of models with very different properties.
The model with ${\cal P}_{vc}^x=+1$
was used in Ref.~\onlinecite{cook-nc17} to describe the band-edge photoconductivity of monolayer GeS, while the model with
${\cal P}_{vc}^x=-1$ applies to BC$_2$N-A2.

(Regarding $M_z$ symmetry, the model {in \eq{H2D}} has ${\cal P}_{vc}^z=+1$ because all atomic orbitals have the same parity and lie on the same plane. Hence, $M_z$ imposes no further constraints.)

Starting from the {two}-band Hamiltonian in \eq{H2D},
the matrix element in \eq{approx-sigma} can be evaluated as described in Refs.~\onlinecite{cook-nc17,yang_divergent_2017}. 
The nonzero components are
\begin{subequations}
\begin{align}
I^{yyy}_{vc}(\omega) &= \frac{8v_y\alpha_y \Delta}{\omega^3},\label{eq:edgexxx}\\
I^{xxy}_{vc}(\omega) &= 
{\frac{(4v_y\alpha_x + 2v_x\alpha_{xy}) \Delta}{\omega^3}}{=I^{xyx}_{vc}(\omega)},\label{eq:edgexxy}\\
I^{yxx}_{vc}(\omega) &=  
\frac{4v_x\alpha_{xy} \Delta}{\omega^3}.
\label{eq:edgeyxx}
\end{align}
\end{subequations}
Using \eq{even-odd} we find that
when $P_{vc}^x=+1$ the $yxx$ component vanishes,
while for $P_{vc}^x=-1$ it is the $yyy$ component that vanishes. These results are in agreement with the first three columns of Table~\ref{table} {for the case ${\cal P}_{vc}^z=+1$}.

Besides illustrating the mirror selection rules, our model reveals a simple quantitative  relation,
\begin{equation}
\label{eq:yxx2xxy}
{\sigma^{yxx} = 2\sigma^{xxy},\quad
\text{when ${\cal P}_{vc}^x=-1$ and ${\cal P}_{vc}^z=+1$},}
\end{equation}
between the two surviving components of the band-edge photoconductivity. 
The above
relation is satisfied rather well by our {\it ab initio} spectrum throughout the entire band-edge region in \fig{2}(a). 
This can be understood from the fact that \eq{yxx2xxy} is quite robust: it follows directly from \eq{Imn}
once we set ${\cal P}_{vc}^x=-1$ and ${\cal P}_{vc}^z=+1$ in \eq{px},
and use the identity $r^{y;x}_{cv}=r^{x;y}_{cv}$ that holds for any two-band tight-binding model 
once off-diagonal position matrix elements are discarded~\cite{PhysRevB.97.245143}.}

\section{Discussion}

To conclude, we discuss the prospects for realizing the physics described herein.
The experimental evidence for the stacking sequence in graphitic BC$_2$N 
remains inconclusive~\cite{PhysRevLett.77.187,PhysRevLett.83.2406}. 
According to DFT {calculations}, the two most stable polytypes 
are the A2 structure studied in this work and a B-type structure denoted B12,
with a difference in formation energy 
of only 1.2~meV/atom favoring the latter~\cite{PhysRevB.73.193304}. 
Both are indirect-gap semiconductors, 
and while B12  provides a slightly better
qualitative match to the experimental
band structure~\cite{PhysRevB.73.193304}, 
neither of them fits {quantitatively} the measured direct and indirect 
gaps~\cite{PhysRevLett.77.187,PhysRevLett.83.2406}. 
Further work
is clearly needed to establish the stacking sequence in bulk BC$_2$N samples, and the linear BPVE could be useful in this regard as it is only present in acentric (A-type) structures.

One intriguing possibility is that it may be possible to grow both polytypes of BC$_2$N using current synthesis techniques.
This has been achieved for other layered materials, such as transition metal dichalcogenides~\cite{TMDReview}. For example, bulk MoS$_2$ 
grows
in two different 
polytypes, centrosymmetric 2H and noncentrosymmetric 3R, 
and the effects of inversion symmetry breaking can be clearly detected in the latter~\cite{Suzuki14}. The reported energy difference between them ranges from~0.1 to 2~meV/atom depending on 
the calculation~\cite{cheng_van_2013,suh_reconfiguring_2018,chen_critical_2013}, which is comparable to that between {the} B12 and A2 {structures of} BC$_2$N~\cite{PhysRevB.73.193304}. We hope that our work  will stimulate similar progress in {graphitic} BC$_2$N, enabling the unambiguous
{identification} of the A2 phase via its large {and highly anisotropic} photogalvanic effect.

\emph{Acknowledgements} --
We thank Stepan S. Tsirkin
for sharing the {\tt irrep} computer code that determines the mirror eigenvalues of the 
Bloch states~\cite{irrep}, and for a previous collaboration on 
related work.
This work was supported by Grant No.~FIS2016-77188-P
from the Spanish Ministerio de Econom\'ia y Competitividad.
This project has received funding from the European Union's Horizon 2020 research and innovation
programme under the Marie Sklodowska-Curie grant agreement No 839237.  

\bibliography{biblio}

\end{document}